\documentclass{Interspeech}
\usepackage{booktabs}
\usepackage{multirow}

\usepackage{algorithm}
\usepackage{algorithmic}

\interspeechcameraready

\title{Cocktail-Party Audio-Visual Speech Recognition}

\author[affiliation={1}]{Thai-Binh}{Nguyen}
\author[affiliation={2}]{Ngoc Quan}{Pham}
\author[affiliation={1,2}]{Alexander}{Waibel}

\affiliation{}{Karlsruhe Institute of Technology}{Germany}
\affiliation{}{Carnegie Mellon University}{USA}
\email{thai-binh.nguyen@kit.edu}

\usepackage{comment}

\begin{document}

\maketitle

\begin{abstract}

Audio-Visual Speech Recognition (AVSR) offers a robust solution for speech recognition in challenging environments, such as cocktail-party scenarios, where relying solely on audio proves insufficient.  However, current AVSR models are often optimized for idealized scenarios with consistently active speakers, overlooking the complexities of real-world settings that include both speaking and silent facial segments. This study addresses this gap by introducing a novel audio-visual cocktail-party dataset designed to benchmark current AVSR systems and highlight the limitations of prior approaches in realistic noisy conditions. Additionally, we contribute a 1526-hour AVSR dataset comprising both talking-face and silent-face segments, enabling significant performance gains in cocktail-party environments. Our approach reduces WER by 67\% relative to the state-of-the-art, reducing WER from 119\% to 39.2\% in extreme noise, without relying on explicit segmentation cues.
  
\end{abstract}

\section{Introduction}
\label{section:introduction}



The visual information obtained from observing a person speak can alter the way auditory signals are perceived, a phenomenon known as the McGurk effect \cite{mcgurk1976hearing}. In cocktail-party environments, even strong ASR models \cite{anwar23_interspeech, nguyen2020super} which mark a significant advance over early efforts \cite{zeppenfeld1997recognition} in conversational speech, still experience significant performance degradation. In such challenging conditions, combining visual cues, like facial movements, with auditory input significantly improves speech comprehension \cite{duchnowski94_icslp, waibel1996multimodal}. Inspired by this interplay, AVSR systems have been developed to leverage visual cues for enhancing speech recognition, particularly in noisy environments. This concept has been explored and validated over the past several decades since its introduction in 1976 \cite{stiefelhagen1999modeling, bub1995knowing, yang1998visual, stiefelhagen2001estimating, waibe112005chil}.

Recent advancements in AVSR have been largely propelled by deep learning models, including the adoption of end-to-end architectures like Transformer \cite{shi22_interspeech} and Conformer \cite{10096889}. These models have been enhanced by improved data utilization, such as pre-training with self-supervised methods like AV-HuBERT \cite{shi22_interspeech}, or by leveraging pre-trained ASR models to generate transcriptions for unlabeled AV datasets, as demonstrated in \cite{10096889, rouditchenko24_interspeech}.  While the combination of audio and visual modalities is expected to make these models robust to noise, our experiments reveal a significant performance decline in cocktail-party environments. For instance, the state-of-the-art (SOTA) AVSR model, Auto-AVSR, achieves an impressive 1.5\% Word Error Rate (WER) on the LRS2 dataset \cite{10096889}, but when background speech noise is added, its WER rises drastically to 69\%. This performance drop is similarly observed in other SOTA models, as discussed in the experiment section, highlighting critical concerns about their practicality in real-world noisy scenarios.


Applying AVSR to the cocktail-party problem is an active area of research. Studies such as \cite{chao16_interspeech, wu21e_interspeech, 10096893} utilize visual information as a query to isolate the target speaker in a mixed audio signal, leveraging the visual modality to focus on and transcribe the speech of a specific individual. Rather than directly outputting the target speaker's transcription, other approaches, like \cite{10.1145/3197517.3201357, gao2021visualvoice, 10447679}, use visual features to extract the target speech signal. A common characteristic of these studies is their reliance on datasets such as LRS2 \cite{8099850}, LRS3 \cite{afouras2018lrs3tedlargescaledatasetvisual}, and VoxCeleb2 \cite{chung18b_interspeech}, which are augmented by mixing utterances to create training and evaluation data with perfect alignment. In these datasets, the speech consistently originates from the speaker shown in the video, or the visual input always depicts a talking face. This alignment allows models to reliably identify the target speaker and generate outputs based on the visible speaker. Recent work, such as \cite{10472123}, has attempted to introduce out-of-sync audio-visual pairs to simulate more challenging scenarios. However, even in these cases, evaluations are typically conducted on datasets like LRS3, which do not reflect the complexity of real-world cocktail-party environments.

A key limitation of commonly used audio-visual datasets like LRS2 and LRS3 is that they fail to capture the complexities of cocktail-party scenarios. Firstly, these datasets predominantly feature single-speaker samples, which do not reflect the multi-speaker interactions typical of cocktail-party environments. Secondly, in noisy environments, a critical challenge is determining whether a visible speaker is actively talking or not. This necessitates the inclusion of both talking face and silent face within an utterance, a factor largely overlooked in current datasets. For the first issue, simulating multi-speaker noise, has been partially addressed through methods such as randomly mixing utterances or adding artificial noise to samples \cite{chao16_interspeech, wu21e_interspeech, 10096893, 10472123}. However, the second issue, involving scenarios with silent face and ambiguous speech activity, remains underexplored. A notable exception is the Chinese multi-channel audio-visual conversation dataset, MISP \cite{9746683}, which attempts to fill this gap. Nevertheless, studies utilizing the MISP dataset \cite{10433931, 10627330} predominantly focus on tasks like audio-visual speaker diarization or audio-visual target speaker extraction (AVTSE) prior to conducting AVSR. While valuable, these approaches address only specific challenges and fall short of tackling the broader set of obstacles faced by AVSR models in realistic cocktail-party environments.

In this study, we focus on developing end-to-end AVSR models tailored to handle cocktail-party scenarios. Our contributions are as follows: (1) we define a novel audio-visual cocktail-party dataset that differs significantly from MISP in three key aspects—it is an English dataset, includes multiple overlapping conversations, and features single-channel audio data. The inclusion of overlapping conversations and single-channel audio makes our dataset more challenging and closer to real-world scenarios compared to MISP, which primarily consists of single conversations per recording and uses multi-channel audio \cite{9746683}. (2) We introduce a 1526-hour AVSR dataset that addresses the limitations of previous datasets by incorporating silent-face utterances, which are crucial for distinguishing active speech. (3) We propose a robust data pipeline for augmenting AVSR datasets, improving their suitability for training models capable of handling cocktail-party environments. (4) Finally, we train a strong baseline AVSR model that demonstrates effective performance on our cocktail-party dataset.


\begin{table*}[h]
\caption{WER (\%) of models on the LRS2 dataset. $\bigstar$ denotes our fine-tuned model. "AV" (in the modality column) indicates models that utilize both audio and visual features.}
\label{tab:lrs2}
\centering
\begin{tabular}{clccccccccc}
\hline
\multirow{2}{*}{Model ID} & \multicolumn{1}{c}{\multirow{2}{*}{Model}} & \multirow{2}{*}{Modality} & \multirow{2}{*}{Train dataset} & \multirow{2}{*}{Interferer} & \multicolumn{5}{c}{SNR (dB)}                                                                                        & \multirow{2}{*}{Avg} \\ \cline{6-10}
                          & \multicolumn{1}{c}{}                       &                           &                           &                             & -5                   & 0                    & 5                    & 10                   & $\infty$                     &                      \\ \hline
\multirow{3}{*}{AV1}      & \multirow{3}{*}{AV-HuBERT CTC/Attention$^\bigstar$}    & \multirow{3}{*}{AV}       & \multirow{3}{*}{lrs2,vox2,avyt}        & 0                           &                      &                      &                      &                      & 2.1                     & \multirow{3}{*}{\textbf{4.1}}    \\
                          &                                            &                           &                           & 1                           & \textbf{6.4}         & \textbf{3.5}         & \textbf{3.4}         & \textbf{2.8}         &                         &                      \\
                          &                                            &                           &                           & 2                           & \textbf{9.0}         & \textbf{4.4}         & \textbf{3.2}         & \textbf{2.8}         &                         &                      \\ \hline
\multirow{3}{*}{AV2}      & \multirow{3}{*}{Conformer CTC/Attention$^\bigstar$}                 & \multirow{3}{*}{AV}       & \multirow{3}{*}{lrs2,vox2,avyt}        & 0                           &                      &                      &                      &                      & 10.9                    & \multirow{3}{*}{17.0}    \\
                          &                                            &                           &                           & 1                           & 19.6                 & 17.1                 & 18.0                 & 15.7                 &                         &                      \\
                          &                                            &                           &                           & 2                           & 20.1                 & 17.6                 & 18.2                 & 16.1                 &                         &                      \\ \hline
\multirow{3}{*}{AV3}      & \multirow{3}{*}{Auto-AVSR \cite{10096889}}                 & \multirow{3}{*}{AV}       & \multirow{3}{*}{\begin{tabular}[c]{@{}c@{}}lrs2,vox2\\lrs3\\avspeech\end{tabular}}         & 0                           &                      &                      &                      &                      & 1.7           & \multirow{3}{*}{21.7}    \\
                          &                                            &                           &                           & 1                           & 56.6                 & 16.6                 & 10.3                 & 4.2                  &                         &                      \\
                          &                                            &                           &                           & 2                           & 69.6                 & 21.8                 & 11.7                 & 3.5                  &                         &                      \\ \hline
\multirow{3}{*}{AV4}      & \multirow{3}{*}{Muavic-EN \cite{anwar23_interspeech}}                 & \multirow{3}{*}{AV}       & \multirow{3}{*}{lrs3}         & 0                           &                      &                      &                      &                      & 7.2                     &          \multirow{3}{*}{12.4}            \\
                          &                                            &                           &                           & 1                           & 18.9                 & 10.8                 & 9.7                  & 8.5                  &                         &                      \\
                          &                                            &                           &                           & 2                           & 25.4                 & 12.1                 & 9.8                  & 8.8                  &                         &                      \\ \hline
\multirow{3}{*}{AV5}      & \multirow{3}{*}{Whisper-Flamingo \cite{rouditchenko24_interspeech}}          & \multirow{3}{*}{AV}       & \multirow{3}{*}{lrs3,vox2}         & 0                           &                      &                      &                      &                      & 6.1                     & \multirow{3}{*}{40.1}    \\
                          &                                            &                           &                           & 1                           & 96.9                 & 37.4                 & 26.2                 & 12.1                 &                         &                      \\
                          &                                            &                           &                           & 2                           & 99.6                 & 38.6                 & 30.6                 & 13.4                 &                         &                      \\ \hline
\multirow{3}{*}{A1}       & \multirow{3}{*}{Whisper large-v3 \cite{radford2022whisper}}          & \multirow{3}{*}{Audio}    & \multirow{3}{*}{5M hours}         & 0                           &                      &                      &                      &                      & 3.7                     & \multirow{3}{*}{33.8}    \\
                          &                                            &                           &                           & 1                           & 97.7                 & 30.9                 & 13.2                 & 6.5                  &                         &                      \\
                          &                                            &                           &                           & 2                           & 99.9                 & 31.1                 & 14.8                 & 6.5                  &                         &                      \\ \hline
\multirow{3}{*}{A2}       & \multirow{3}{*}{Auto-AVSR \cite{10096889}}                 & \multirow{3}{*}{Audio}    & \multirow{3}{*}{same as AV3}         & 0                           & \multicolumn{1}{l}{} & \multicolumn{1}{l}{} & \multicolumn{1}{l}{} & \multicolumn{1}{l}{} & \textbf{1.5} & \multirow{3}{*}{34.7}    \\
                          &                                            &                           &                           & 1                           & \multicolumn{1}{l}{93.9} & \multicolumn{1}{l}{30.5} & \multicolumn{1}{l}{22.7} & \multicolumn{1}{l}{5.3} & \multicolumn{1}{l}{}    &                      \\
                          &                                            &                           &                           & 2                           & \multicolumn{1}{l}{95.8} & \multicolumn{1}{l}{33.0} & \multicolumn{1}{l}{23.7} & \multicolumn{1}{l}{6.2} & \multicolumn{1}{l}{}    &                      \\ \hline
V1                        & Auto-AVSR \cite{10096889}                                  & Visual                    &        same as AV3                   & 0                           & \multicolumn{5}{c}{15.7}  &                                                                                 \multicolumn{1}{c}{15.7}                                 \\ \hline

\end{tabular}
\vspace{-1.5em}
\end{table*}

\begin{figure}[t]
  \centering
  \includegraphics[width=0.6\linewidth]{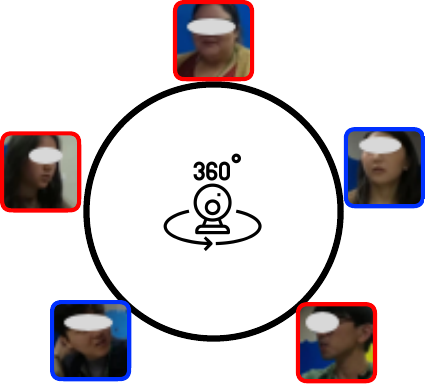}
  \caption{Schematic overview of AVCocktail's recording scene.}
  \label{fig:avcocktail_setting}
  \vspace{-2em}
\end{figure}

\vspace{-0.5em}
\section{Cocktail-Party AVSR}

\subsection{Task definition}

Given an input sequence of audio $ A = \{a_1, a_2, \dots, a_T\} $ and video $ V = \{v_1, v_2, \dots, v_T\} $, where $T$ is the total number of time steps, $a_t$ represents the audio feature and $v_t$ represents the visual feature at time step $t$. The task is to predict the transcription $Y_{\text{target}} = \text{AVSR}(A, V) = \{y_1, y_2, \dots, y_N\} $, where $N$ is the length of the target speech transcription, $\text{AVSR}$ denotes the model that takes both audio and visual features as input to generate the transcription. While the video captures the target speaker, the audio may include overlapping speech, background noise, and uncertainty in speech activity when the target speaker is visible but not speaking.

\vspace{-0.5em}
\subsection{Baseline}
\label{subsection:baseline}

We adopt two off-the-shelf architectures to evaluate the effectiveness of the proposed method through fine-tuning. The first model, AV-HuBERT CTC/Attention (AV1), uses AV-HuBERT \cite{shi22_interspeech} as the encoder and the decoder integrates a projection layer and a Transformer decoder with joint CTC/Attention training \cite{8068205}. The second model is the Conformer CTC/Attention architecture (AV2) proposed by \cite{9414567}, where the encoder consists of two Conformer blocks: one for audio and one for visual feature extraction. The decoder is identical to AV1, employing joint CTC/Attention training.

In addition to fine-tuning the above architectures, we directly evaluate recent AVSR models that have achieved SOTA performance on the LRS2 and LRS3 datasets as strong baselines. These include Auto-AVSR (denoted as AV3) \cite{10096889}, which uses a Conformer with CTC/Attention and is trained on 3448 hours of AVSR data. Two additional variants of AV3 are included for benchmarking: a visual-only model (V1) and an audio-only model (A2). Another baseline is the Muavic-EN (AV4) model \cite{anwar23_interspeech}, which employs AV-HuBERT as the encoder and a Transformer decoder, trained exclusively on the LRS3 dataset with various types of additive noise. Whisper-Flamingo (AV5) \cite{rouditchenko24_interspeech} combines AV-HuBERT with Whisper-large and is trained on LRS3, Vox2, and augmented noisy data. Lastly, Whisper-large (A1), trained on 5M hours of diverse audio data, serves as a strong baseline for benchmarking audio-only performance. 

\vspace{-0.5em}
\section{Data Preparation}

In this study, we utilize four datasets. For training, we use LRS2 (train and pretrain sets), Vox2 (train set), and AVYT. For testing, we evaluate on the LRS2 test set (including a modified version) and AVCocktail. Details of LRS2 test set, AVYT, and AVCocktail are provided in the following subsections. About Vox2, we simple employ Whisper-large \cite{anwar23_interspeech} to transcribe the audio and retain only the English segments.

\subsection{Lip Reading Sentences 2 (LRS2)}
\label{subsection:lrs2}

LRS2 \cite{8099850} is a widely used AVSR dataset commonly utilized for benchmarking\footnote{LRS3 was unavailable at the time of this study}. To evaluate the performance of SOTA AVSR models in cocktail-party settings, we augment LRS2 by randomly adding interfering speakers in the background with varying signal-to-noise ratios (SNR). Specifically, we introduce up to two interfering speakers, with SNR levels controlled at \(-5\), \(0\), \(5\), and \(10\) dB. The original LRS2 dataset, without any interfering speakers, corresponds to an SNR of \(\infty\) and zero interferer.

\subsection{AVCocktail}

In our AVCocktail dataset, we focus on a scenario where people gather around a table, forming small groups of 2 to 5 individuals, each group engaged in a discussion topic. Figure \ref{fig:avcocktail_setting} illustrates a recording scene with two such groups. A single 360-degree camera is placed at the center of the table to capture all participants' faces. From the 360-degree footage, we extract $224 \times 224$ cropped video clips and a single audio channel for each speaker. Each recording session lasts 5 to 7 minutes, resulting in a total evaluation set of approximately 6.1 hours of video from 45 speakers. All individual video then been segmented and transcribe by human.

\subsection{Automatic AVSR dataset from Youtube (AVYT)}

\begin{figure}[t]
  \centering
  \includegraphics[width=1.0\linewidth]{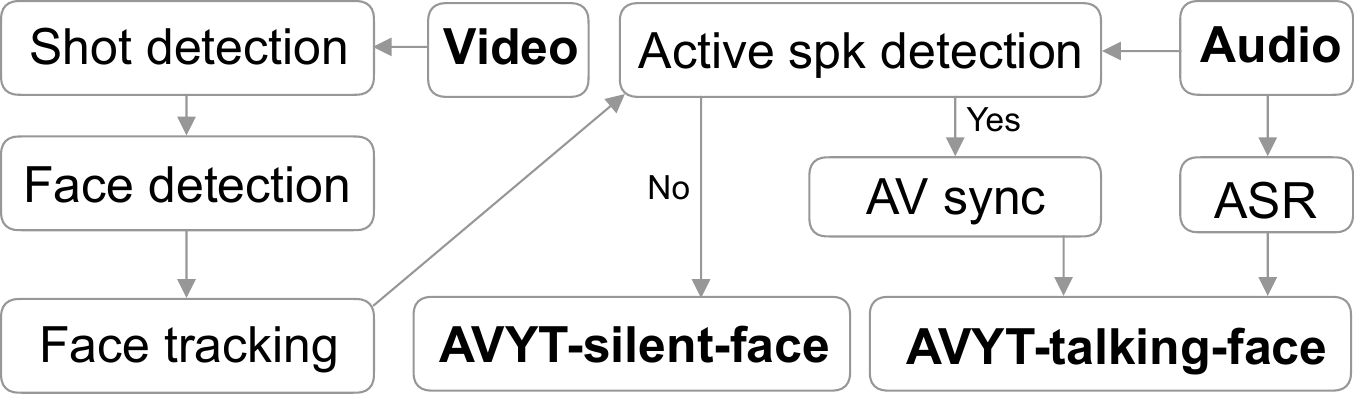}
  \caption{Pipeline to generate AVYT dataset}
  \label{fig:avyt_pipeline}
  \vspace{-2em}
\end{figure}

Due to the limitations of existing AVSR datasets, as described in Section \ref{section:introduction}, we identified the need for an additional dataset to better train AVSR models for cocktail-party scenarios. To address this, we introduce AVYT, derived from the 6,533 hours of YouTube content introduced in \cite{retkowski-waibel-2024-text}. AVYT consists of two subsets: a silent-face set with 77 hours spanning 79k clips, and a talking-face set containing 1449 hours across 666k clips. 

Figure \ref{fig:avyt_pipeline} illustrates the processing pipeline for constructing the AVYT dataset. The first step, inspired by the LRS2 data processing pipeline \cite{8099850}, involves shot detection, face detection, and face tracking. This process yields cropped video clips containing a single speaker. In the second step, an active speaker detection model \cite{Liao_2023_CVPR} identifies talking-face segments and silent-face segments, assigning them to separate sets. The third step further refines the talking-face clips: AV sync \cite{Chung2016OutOT} ensures proper audio-video synchronization, and the synchronized clips are then transcribed using Whisper-large \cite{anwar23_interspeech} to create the final talking-face set.

\subsection{Data augmentation pipeline}
\label{subsection:data_augment_pipeline}
\vspace{-1em}
\begin{algorithm}[H]
\caption{Data Augmentation Pipeline}
\label{algorithm:data_augmentation_pipeline}
\begin{algorithmic}[1]
\STATE \textbf{Function} generate\_sample(dialog, silent\_face, interferer)
\FOR{video in [Vox2, LRS2, AVYT-talking-face]}
    \STATE videos = [video]
    \IF{dialog}
        \STATE videos += n\_rand([Vox2, LRS2, AVYT-talking-face])
        \IF{silent\_face}
            \STATE videos += n\_rand(AVYT-silent-face)
        \ENDIF
    \ENDIF
    \STATE avsr\_sample = concat(shuffle(videos))
    \IF{interferer}
        \STATE avsr\_sample = augment\_speech\_noise(avsr\_sample)
    \ENDIF
    \STATE avsr\_sample = video\_transform(avsr\_sample)
    \STATE \textbf{yield} avsr\_sample
\ENDFOR
\end{algorithmic}
\end{algorithm}
\vspace{-1em}

Algorithm \ref{algorithm:data_augmentation_pipeline} outlines the data augmentation pipeline designed to construct samples for fine-tuning our AVSR model. The pipeline incorporates three key augmentation strategies: dialog augmentation, adding silent-face video clips, and introducing interfering speakers. These strategies can be applied individually or combined, with the option to enable or disable each one as needed. Dialog augmentation merges multiple video clips to simulate conversational scenarios, a technique proven effective in prior studies \cite{10446589, 10889116}. This approach is particularly crucial when combined with the second augmentation strategy, which integrates silent-face video clips. Since silent-face clips are labeled with $<unk>$ transcripts, merging multiple clips into a dialog-like structure prevents the model from relying solely on visual cues (frame sequence differences) to infer transcripts for silent-face segments. The third strategy introduces interferer speakers in the background at varying SNR. Finally, the pipeline applies video transformations, including horizontal flipping, random cropping, and adaptive time masking for both visual and audio streams, to further diversify the training data. The video frames are cropped to the mouth region of interest (ROI) using a $96\times96$ bounding box, while the audio is sampled at a 16 kHz rate, similar to \cite{shi22_interspeech, 9414567}.

\vspace{-0.5em}
\section{Experimental setup}

As described in Section \ref{subsection:baseline}, we fine-tune two model architectures. The AV-HuBERT CTC/Attention (AV1) model uses the AV-HuBERT large \cite{shi22_interspeech} as the encoder, which has 24 transformer blocks, each with 16 attention heads. The CTC/Attention decoder is a 6-layer Transformer with the same dimensions and number of attention heads as the encoder. The Conformer CTC/Attention (AV2) \cite{9414567} consists of 12 encoder layers for both audio and visual inputs, each with 16 attention heads. The decoder CTC/Attention in this model is similar to the first, consisting of a 6-layer Transformer.

The data pre-processing pipeline used to train our model is detailed in Section \ref{subsection:data_augment_pipeline}. We then conduct several evaluations. First, we compare the performance of our fine-tuned model with current SOTA AVSR models on the LRS2 test set (including a modified version), as described in Section \ref{subsection:lrs2}. This benchmark evaluates how model performance degrades under data distortions, similar to the challenges posed by cocktail-party scenarios. The second experiment assesses these models on the real cocktail-party dataset, AVCocktail. Since each recording lasts 5 to 7 minutes, a segmentation step is required before inference. We employ three segmentation strategies. (1) a model-based approach using Active Speaker Detection \cite{Liao_2023_CVPR}, which leverages both audio and visual cues to determine speaking segments. (2) a fixed 10-second chunk-based approach with a sliding window and no overlap and (3) manual segmentation, where segments are labeled by humans. Finally, we conduct an ablation study on the data augmentation techniques introduced in Section \ref{subsection:data_augment_pipeline} to identify the most influential factors in AVSR model performance. 


\begin{table}[]
\caption{WER (\%) of models on the AVCocktail dataset}
\label{tab:avcocktail}
\centering
\begin{tabular}{cccc}
\hline
\multirow{2}{*}{\begin{tabular}[c]{@{}c@{}}Recognition\\  Model \\ID\end{tabular}} & \multicolumn{3}{c}{Segmentation}                                                                                                                    \\ \cline{2-4} 
                       & \begin{tabular}[c]{@{}c@{}}Active  Speaker \\  Detection \cite{Liao_2023_CVPR}\end{tabular} & \begin{tabular}[c]{@{}c@{}}Fixed \\ chunk (10s)\end{tabular} & Gold          \\ \hline
AV1                    & \textbf{22.6}                                                        & \textbf{39.2}                                                & \textbf{18.2} \\
AV2                    & 48.4                                                                 & 89.5                                                         &        41.9       \\
AV3                    & 74.6                                                                 & 133.2                                                        &        67.8       \\
AV4                    & 35.6                                                                 & 119.0                                                        & 26.1          \\
AV5                    & 70 .8                                                                 & 133.3                                                        & 58.3          \\
A1                     & 67.4                                                                 & 143.9                                                        & 54.7          \\
A2                     & 75.8                                                                 & 131.7                                                        &       70.3        \\
V1                     & 56.0                                                                 & 167.9                                                        &       49.9        \\ \hline

\end{tabular}
\vspace{-1.5em}
\end{table}

\vspace{-1em}
\section{Results}

Table \ref{tab:lrs2} presents the WER (\%) of baseline models and our fine-tuned models on the LRS2 test set, including both the original and modified versions. The WERs for models evaluated on the original LRS2 test set are shown in the column where SNR = $\infty$. Overall, all models perform well on the original clean LRS2 dataset. The best-performing model in this setting is A2, the Conformer CTC/Attention audio-only model, which is expected since the dataset consists of clean speech, and A2 is well-trained with in-domain data.

However, performance degrades significantly as SNR decreases and the number of interfering speakers increases. Audio-only models are the most affected by noise, with A1’s WER rising from 3.7\% to 99.9\% and A2’s WER increasing from 1.5\% to 95.8\%. In contrast, the Conformer CTC/Attention visual-only model (V1) remains unaffected by noise, maintaining a constant WER of 15.7\%. Among the baseline audio-visual models (AV3, AV4, AV5), despite leveraging visual features and noise augmentation during training, performance still deteriorates significantly under noisy conditions. Notably, AV3 and AV5 suffer the most, with WERs rising from 1.7\% and 6.1\% to 69.6\% and 99.6\%, respectively. AV4 demonstrates the highest noise robustness among the baselines, likely due to its augmentation with both speech noise and additional noise types beyond "natural," "music," and "babble," which were used in AV3. Although AV5 employs the same augmentation strategy as AV4, it appears to rely more on audio than visual information, leading to the worst performance under extreme noise conditions.

AV1 and AV2 are our fine-tuned models, trained with in-domain data as described in Section \ref{subsection:data_augment_pipeline}. AV2 is initialized from AV3’s parameters. After fine-tuning, AV2 achieves overall better performance than AV3 (17.0\% WER vs. 21.7\%), but its WER on the original LRS2 test set increases significantly from 1.7\% to 10.9\%. AV1, which employs AV-HuBERT as the encoder and a CTC/Attention decoder, achieves the best performance among all models, with a WER of 2.1\% on the original LRS2 test set and an average WER of 4.1\% overall.

Unlike LRS2, AVCocktail consists of long video recordings that contain both talking-face and silent-face segments. The choice of segmentation method influences the proportion of silent-face segments included in the inference data. In the AVCocktail dataset, the total speaking duration accounts for 46.3\% of the recordings. Active Speaker Detection (ASD) achieves a precision of 84.4\%, recall of 97.8\%, and F1-score of 90.6\% in detecting speaking segments. This means that with ASD-based segmentation, nearly all talking-face segments are retained, though some silent-face segments are incorrectly detected as speech. In contrast, fixed-length segmentation inherently includes both talking-face (46.3\%) and silent-face (53.7\%) segments, as it processes the video in uniform chunks without considering whether the target speaker is speaking or silent.

Table \ref{tab:avcocktail} presents the performance of different models on the AVCocktail dataset. In general, a higher proportion of silent-face segments leads to worse WER. For baseline models (AV[3-5], A[1-2], and V1), the WER in the extreme case of fixed-length segmentation exceeds 100\%, while our AV1 model still achieves a WER of 39.2\% in this scenario. When using a segmentation model like ASD, the WER improves significantly. As expected, the best performance is achieved with gold segmentation. Our AV1 model demonstrates strong robustness, achieving the best performance among all baseline models across all types of segmentation.

Table \ref{tab:avcocktail_abbv} presents the ablation study on the impact of different data factors on AV-HuBERT CTC/Attention performance in the AVCocktail dataset, using ASD for segmentation. A total of six experiments were conducted. Training with only conventional AVSR datasets (LRS2 and Vox2) results in a WER of 58.8\%. Adding speech noise improves performance by 44.7\% relative (WER reduced to 32.5\%). Incorporating the AVYT-talking-face dataset further reduced the WER by 7.1\%, reaching 30.2\%. Dialog augmentation alone had a minimal impact, slightly decreasing the WER to 29.4\%. Using full AVYT (both talking-face and silent-face sets) without dialog augmentation led to a WER of 28.5\%. The most substantial improvement was achieved by combining dialog augmentation with the full AVYT dataset, achieving a significantly lower WER of 22.6\%.

\begin{table}[]
\caption{Ablation study on the impact of data factors on AV-HuBERT CTC/Attention performance in AVCocktail dataset.}
\label{tab:avcocktail_abbv}
\centering
\begin{tabular}{lccccc}
\hline
Dataset        & Interferer & Dialog & Silent-face &  WER  \\ \hline
lrs2,vox2      &        &         &    & 58.8 \\
lrs2,vox2      &  \checkmark      &         &    & 32.5 \\
$+$AVYT-talking & \checkmark       &         &    &  30.2\\
$+$AVYT-talking & \checkmark      &   \checkmark      &    & 29.4 \\
$+$AVYT & \checkmark       &         &  \checkmark  &  28.5\\
$+$AVYT & \checkmark      & \checkmark       &  \checkmark  & 22.6 \\ \hline
\end{tabular}
\vspace{-1.5em}
\end{table}

\vspace{-0.5em}
\section{Conclusion}

In this study, we benchmarked SOTA AVSR models, which perform impressively on conventional datasets like LRS2/LRS3 but struggle with cocktail-party scenarios. We highlighted the gap between conventional datasets and real-world cocktail-party scenarios, where target speakers are not always active. The presence of silent-face segments significantly impacts AVSR model performance, as the model tends to hallucinate output. To address this gap, we introduced the AVYT dataset and a data augmentation pipeline to improve model robustness. Additionally, we created AVCocktail, the first English audio-visual cocktail-party benchmark, to evaluate AVSR performance in realistic multi-speaker settings. All datasets and models are publicly available for further research at: https://github.com/nguyenvulebinh/AVSRCocktail

\section{Acknowledgment}
\vspace{-0.2em}
The authors gratefully acknowledge support from Carl Zeiss Stiftung under the project Jung bleiben mit Robotern (P2019-01-002). This work was also partially supported by the European Union’s Horizon research and innovation programme (grant No. 101135798, project Meetween), the Volkswagen Foundation project ``How is AI Changing Science? Research in the Era of Learning Algorithms'' (HiAICS), and KIT Campus Transfer GmbH (KCT) staff in accordance to the collaboration with Carnegie-AI, as well as the HoreKa supercomputer funded by the Ministry of Science, Research and the Arts Baden-Württemberg and BMBF. Part of this research was supported by a grant from Zoom Video Communications.
\vspace{-1em}
\bibliographystyle{IEEEtran}
\bibliography{mybib}

\end{document}